\documentclass[manuscript]{aastex}

\slugcomment{Not to appear in Nonlearned J., 45.}

\shorttitle{Compact Stars}
 \shortauthors{H. Rodrigues et al.}

\begin{document}

\title{Massive Compact Stars as Quark Stars}

\author{Hil\'ario Rodrigues}
\affil{Centro Federal de Educa\c{c}\~ao Tecnol\'ogica do Rio de
Janeiro \\  Av Maracan\~a 249, 20271-110, Rio de Janeiro, RJ,
Brazil} \email{harg@cefet-rj.br}

\author{S\'ergio Barbosa Duarte}
\affil{Centro Brasileiro de Pesquisas F\'isicas \\ Rua Dr. Xavier
Sigaud 150, 22290-180, Rio de Janeiro, RJ, Brazil}
\email{sbd@cbpf.br}

\and

\author{Jos\'e Carlos T. de Oliveira}
\affil{    Departamento de F\'isica, Universidade Federal de
Roraima \\ Campus do Paricarana s/n, 69310-270, Boa Vista, RR,
Brazil}
 \email{jcto@cbpf.br}

\begin{abstract}
High massive compact stars have been reported recently in the
literature, providing strong constraints on the properties of the
ultradense matter beyond the saturation nuclear density. In view
of these results, the calculations of quark star or hybrid star
equilibrium structure must be compatible with the provided
observational data. But, since the used equations of state
describing quark matter are in general too soft, in comparison
with the equation of states used to describe the hadronic or
nuclear matter, the calculated quark star models presented in the
literature are in general not suitable to explain the stability of
high compact massive objects.

In this work, we present the calculations of spherically symmetric
quark star structure by using an equation of state which takes
into account the superconducting Color-Flavor Locked (CFL) phase
of the strange quark matter. In addition, some fundamental aspects
of QCD (asymptotic freedom and confinement) are considered by
means of a phenomenological description of the deconfined quark
phase, the density-dependent quark mass model
\citep{Chakra1,Chakra2}. The quark matter behavior introduced by
these model stiffens the corresponding equation of state. We thus
investigate the influence of these models on the mass-radius
diagram of quark stars. We obtain massive quark stars due to the
stiffness of the equation of state, when a reasonable
parametrization of the color superconducting gap is used. Models
of quark stars enveloped by a nucleonic crust composed of a
nuclear lattice embedded in an electron gas, with nuclei close to
neutron drip line, are also discussed.

\end{abstract}





\section{Introduction}

The study of the structure of compact stars requires the
understanding of the equation of state describing the stellar
matter under extreme conditions. It has be pointed out that most
of the current equations of state describing quark matter are too
soft and so unable to explain the existence of massive neutron
stars \citep{Cottam,Ozel1}. Only stiff equations of state
describing normal nuclear matter at high densities would be
capable of explaining the stability of high compact star masses
($M \sim 2 \, M_{\sun}$). Thus, apparently these observational
data tend to favor the existence of matter without deconfined
quarks in the interior of neutron stars. Examples of massive
compact stars are the isolated neutron star RX J1856.5-3754
\citep{Trumper}, some low-mass X-ray binaries, e.g., 4U 1636-536
\citep{Barret}, and the neutron star J0751+1807 ($2.1 \pm 0.2 \,
M_{\sun}$) \citep{Nice}.

More recently, two possible pairs of values of mass and radius
have been attributed to neutron star EXO 1745-248, one of them
centered around $M = 1.4 \, M_{\odot}$ with radius $R = 11 \, km$,
and the other one centered around $M = 1.7  \, M_{\sun}$ and with
a smaller radius $R = 9 \, km$ \citep{Ozel2}. The neutron star in
the low-mass X-ray binary 4U 1608-52 has a mass determined as $M
\geq (1.84 \pm 0.009) M_{\sun}$ and radius $R \geq (9.83 \pm
1.24)\ km$ \citep{Ozel3}. Regarding the neutron star EXO
$0748-676$, the obtained lower limits on the mass and radius are:
$ M \geq 2.10\pm 0.28\ M_{\odot } $ and $R\geq 13.8 \pm 1.8 \ km $
\citep{Ozel1}, even though these results are still unclear
\citep{Cottam}.

 It is believed that at high densities the strange
quark matter is a more stable configuration than the ordinary
nuclear matter \citep{Bodmer,Witten} and hence it is claimed that
neutron stars would be stellar objects entirely or partially
composed of strange quark matter \citep{Itoh,Fridolin}, that is, a
quark star or a hybrid star. However, some used equations of state
describing quark matter are too soft to support quark stars with
large masses and then, in this scenario, quark stars or hybrid
stars seem to be incompatible with the observed massive neutron
stars mentioned above.  In fact, calculations using soft equations
of state for quark matter provide values of maximum masses for
hybrid neutron stars around $1.6 \, M_{\sun}$, and lower values of
maximum masses for strange quark stars.

Stiff equations of state of quark matter can be obtained when
effects of strong interaction are taken into account
\citep{Alford1}. Calculations of compact star models using quark
matter equation of state, generated by modified MIT bag model
(including perturbative corrections to QCD) or by the
Nambu-Jona-Lasinio model, are in general capable of reproducing a
maximum star mass around $2 \, M_{\odot}$, which is compatible
with the observational data \citep{Alford2,Rodrigues}.

  A recent calculation for cold
and dense QCD strange quark matter including corrections to order
$O( \alpha_s^2) $ indicates that massive compact stars with mass $
\gtrsim 2 M_{\sun} $ up to maximal masses $ \sim 2.75 M_{\sun} $
would be interpreted as possible candidates for strange quark
stars,  $i. e.$, compact stars composed entirely of deconfined
$u$, $d$ and $s$ quarks, and that compact stars with observed
masses up to $ \sim 2 M_{\sun} $ would be identified with hybrid
stars, neutron stars with a central core made up of deconfined
quark matter \citep{Kurkela}. Motivated by these recent
investigations, in this work we study the color superconducting
quark matter by using a phenomenological model, namely the
density-dependent quark mass model \citep{Chakra1,Chakra2}, in
order to describe the strange quark star structure. Assuming the
CFL phase is the ground state of strange quark matter, we consider
the effects of the CFL gap energy on the global strange quark star
structure.

The density-dependent quark mass model provides stiff equations of
state, and hence large quark star masses can be obtained with this
type of equations of state, which are compatible with some of
those observational data, if reasonable values for the equation of
state parameters are used. Note that these results contrast with
the mass-radius relationships predicted by equations of state
constructed with descriptions of quark matter based on a simple
version of the MIT bag model, as discussed in Ref.
\citep{Oliveira}.

 The study of the structure of the nucleonic crusts is also presented, and
our results for quark star mass versus radius are compared with
the data presented recently by \"{O}zel in references
\citep{Ozel1}, \citep{Ozel2} and \citep{Ozel3}.

This work is organized as follows. In Section 2 we describe the
equation of state of the cold color superconducting strange quark
matter. In Section 3 we study the equation of state of the
unpaired quark matter (UQM) and discuss the color superconducting
to unpaired quark matter phases. In Section 4 we present and
discuss our results obtained for bare quark stars structure and
compare them to some values of mass and radii derived from
observational data. In Section 5 we present some results for quark
stars with nucleonic crusts. Conclusions and final remarks are
presented in Section 6.

\begin{table}
\centering \caption{Set of the used model parameters. The
density-dependent mass model parameters $C$ and $m_{s0}$ are in
units of $MeV . fm^{-3}$ and $MeV$, respectively. The free
characteristic gap parameter $\mu^{\ast}$ is in units of $MeV$.
}\label{t1}
 \vspace{0.2 cm}
\begin{tabular}{|c|c|c|c|c|}
  \hline
  $\mu^{\ast}$ & $C=100$, $m_{s0}=50$  & $C=90$, $m_{s0}=80$  &$C=80$, $m_{s0}=100$  & $C=70$, $m_{s0}=150$ \\
  \hline
  100 & A1 & B1 & C1 & D1 \\
  \hline
  150 & A2 & B2 & C2 & D2 \\
  \hline
  200 & A3 & B3 & C3 & D3 \\
  \hline
  250 & A4 & B4 & C4 & D4 \\
  \hline
\end{tabular}
\end{table}

\section{Description of the cold color superconducting quark matter}

In the last few years the study of color superconducting phase in
quark-gluon plasma \citep{Alford3,Alford4,Alford5} has been
attracting a great interest in discussing the possible states of
quark matter. At QCD perturbative regime the attractive quark
interaction introduces instabilities in the Fermi surface,
producing a gap in the quasiparticle energy spectrum. The color
and flavor symmetries of three-flavor QCD are hence broken down,
leading to the formation of pairs of quarks (this mechanism is
analogous to the electron pairing in the ordinary electric
superconducting phenomenon). The quark matter becomes a color
superconductor, with equal number of quarks for the three flavors
$u$, $d$ and $s$ in an electrically neutral phase called
Color-Flavor Locked (CFL) \citep{Alford6,Muzinger}. The color
superconducting gap is included in an additional energy term to
the thermodynamic potential associated to a quasi-free quark
system as it is done elsewhere (see for example the review article
\citep{Schmitt}).

In order to describe the Color-Flavor Locked phase composed of
free quarks $u$, $d$ and $s$ at zero temperature we use the
thermodynamic potential density given by \citep{Rajagopal}
\begin{equation}
\Omega=\sum_{f}\frac{3}{\pi^{2}}\int_{0}^{\nu}k^{2}(\sqrt{k^{2}
+m_{f}^{2}}-\mu)dk+\Omega_{CFL}  \label{potential} ,
\end{equation}
where $m_f$ stand for the constituent quark masses, with
$f=\{u,d,s\}$. In this equation, the first term in the right-hand
side gives the contribution of degenerate quarks to the
thermodynamic potential density, and the second one is the first
order contribution from the quasiparticle energy gap, which reads
\begin{equation}
\Omega_{CFL}=-\frac{3}{\pi^{2}} \Delta_{CFL}^{2} \mu^{2},
\label{gapCFL}
\end{equation}
where the quark chemical potential and the common Fermi momentum
$\nu$, considering $ m_u = m_d$, are given respectively by
\begin{equation}
\mu=\frac{\mu_{u} + \mu_{d} + \mu_{s}}{3},
\end{equation}
and
\begin{equation}
\nu =\left[ \left( 2\mu
-\sqrt{\mu^{2}+\frac{m_{s}^{2}-m_{u}^{2}}{3}} \right)
^{2}-m_{u}^{2}\right] ^{1/2} \label{nuFermi}  .
\end{equation}

The above common Fermi momentum of the quark system depends on the
mass of the three light quark flavors, and it does minimize the
thermodynamic potential given by Eq. (\ref{potential}) in respect
to the parameter $\nu$. Note that for $m_{u}=m_{d}=0$, one
recovers the expression
\begin{equation}
\nu =2\mu -\sqrt{\mu^{2}+\frac{m_{s}^{2}}{3}},
\end{equation}
derived e.g. by Rajagopal and Wilczek \citep{Rajagopal}, who have
considered massless up and down quarks, and the nonzero current
mass of the strange quark in the calculations.

In the $CFL$ phase described here, the three flavors of quarks
satisfy the following conditions: ($i$) they have equal Fermi
momenta, which minimizes the free energy of the system
\citep{Rajagopal,Orsaria}; ($ii$) they have equal number densities
$n_{f}$, as a consequence of the first condition, which means that
$n_f = \rho_B$ and $\mu_f=\mu$, for $f=\{u,d,s\}$; ($iii$) the
quark system has neutral electromagnetic charge in bulk, as a
direct consequence of the previous condition.

In the density-dependent quark mass model \citep{Chakra1,Chakra2}
the masses of the three lightest quarks scale inversely with the
baryon number density, namely:
\begin{equation}
m_{u} = m_{d} = \frac{C}{3\rho_B} \; \; \; \; \;   {\rm and} \; \;
\; \; \;  m_{s}= m_{s0}+\frac{C}{3\rho_B}  \label{1},
\end{equation}
where $C$ is interpreted as the constant energy density in the
zero quark density limit, and is analogous to the QCD vacuum
energy density present in the MIT bag model, and
\begin{equation}
\rho_B = \frac {n_{u} + n_{d} + n_{s}}{3} \label{2},
\end{equation}
is the baryon number density expressed in terms of the quark
number densities $n_{u}$, $n_{d}$ and $n_{s}$. The current mass of
the strange quark, $m_{s0}$, enters as a parameter of the model.
The dependence of the quark masses on the baryon density mimics
the quark interaction at different values of density,
incorporating the effect of quark interactions in the model.
Througt this artifact, the density-dependent quark mass model
recovers the asymptotic behavior of quark matter predicted by QCD
at high densities, namely, the quark asymptotic freedom and the
dynamical confinement of quarks for low density regime as natural
limit situations. Consequently, in this model quarks are
dynamically massive, and thus chiral symmetry of the QCD
Lagrangian is dynamically broken. Besides, as mentioned
previously, it would be unrealistic adopting vanishing dynamical
quark masses to describe quark matter for the range of densities
currently accepted for the interior of compact stars
\citep{Schmitt}.

The pressure derived from the thermodynamic potential in equation
(\ref{potential}) thus reads
\begin{equation}
P=-\Omega + B^{*}.
\end{equation}

The term $B^{*}$ in the right-hand side of the last equation
arises from the thermodynamic consistence between the degenerate
quark gas pressure and the thermodynamic potential, due to the
dependence of the quark mass on the baryon density. The additional
term $B^{*}$ is similar to QCD vacuum pressure present in the MIT
bag model. The presence of this term enables a smooth model
transition from the quark confinement regime to the asymptotic
freedom since its explicitly form,
\begin{equation}
B^{*}=\rho_B\left.  \frac{\partial\Omega}{\partial
\rho_B}\right\vert _{T=0,\mu },
\end{equation}
is a decreasing function of the baryon density.

We thus found the following explicit expression for the pressure:
\begin{equation}
P=\sum_f\frac{3}{8\pi ^{2}}m_{f}^{4}\left[
f(x)+\frac{4}{m_{f}}\frac{C}{3\rho_B} g(x)\right] +\frac{3}{\pi
^{2}}(\nu ^{3}\mu  + \Delta_{CFL}^{2} \mu^{2}),
\end{equation}
where the functions $f(x)$ and $g(x)$ are given by
\begin{equation}
f(x)=\ln \left[ x+ (x^{2}+1)^{1/2}\right]
-x(x^{2}+1)^{1/2}(2x^{2}+1),
\end{equation}
and
\begin{equation}
g(x)=\ln \left[ x+(x^{2}+1)^{1/2}\right] -x(x^{2}+1)^{1/2},
\end{equation}
with $x$ defined by
\begin{equation}
x=\frac{\nu}{m_{f}}.
\end{equation}

Finally, the energy density can be obtained from the thermodynamic
relation
\begin{equation}
\varepsilon= -P + \sum_{f}n_{f}\mu_{f},
\end{equation}
from which we find
\begin{equation}
\varepsilon_{CFL} = -P + 3 \rho_{B} \mu .
\end{equation}

In order to complete the description of the CFL quark matter
phase, we write the quark number density,
\begin{equation}
n_{f}= -\left.
\frac{\partial\Omega}{\partial\mu}\right\vert_{\rho_B },
\end{equation}
which gives the baryon number density
\begin{equation}
\rho_{B}= \frac{1}{\pi^{2}}(\nu^{3}+2\Delta_{CFL}^{2}\mu)
\label{32} .
\end{equation}

\begin{figure}
\epsscale{.50} \plotone{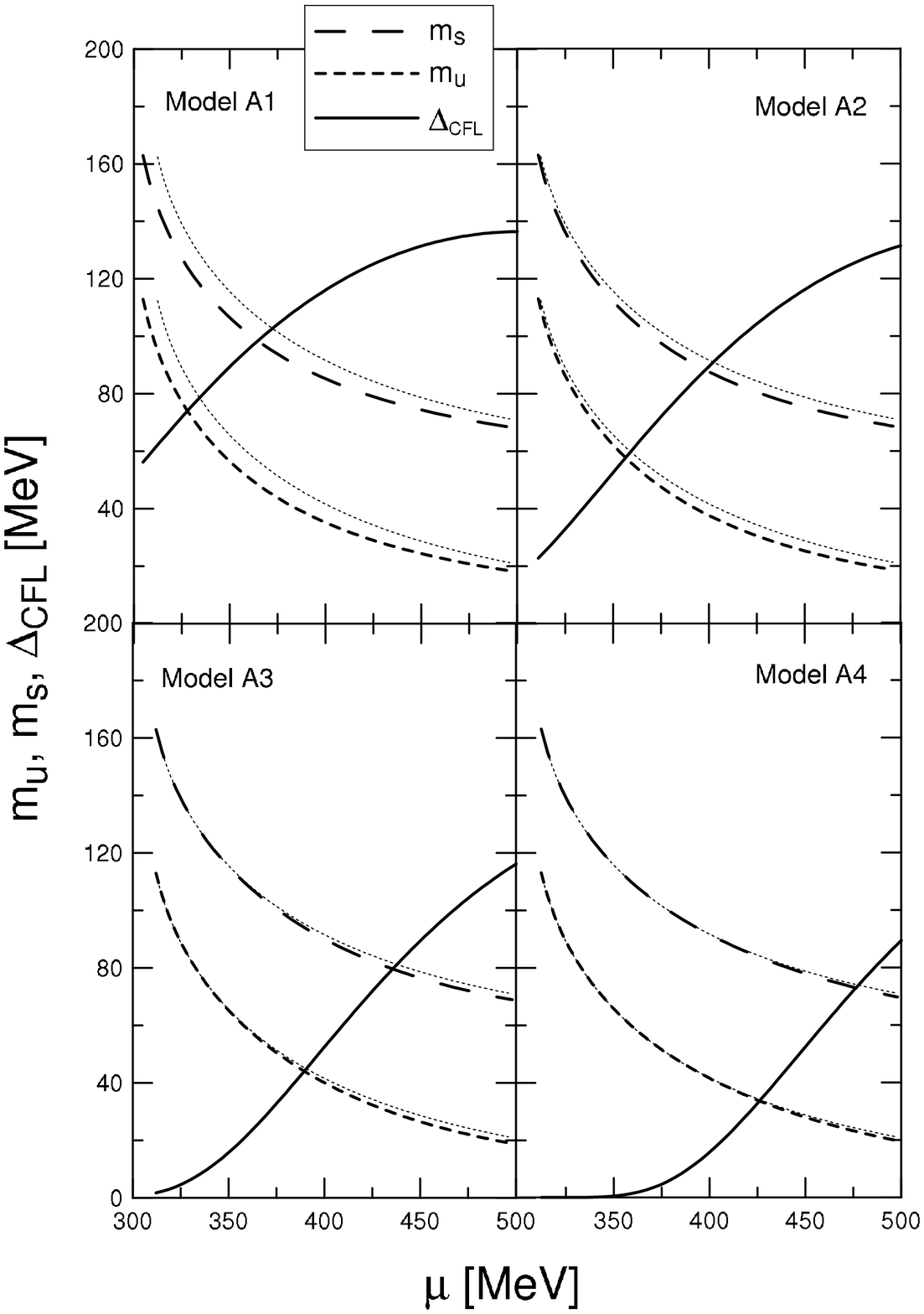} \caption{ Dependence of the
quark masses and the color superconducting gap on the quark
chemical potential for the models A1, A2, A3, and A4 given in
Table \ref{t1}. The thin dotted curves near the constituent quark
masses curves represent the corresponding constituent quark masses
for the unpaired quark phase. \label{fig1}}
\end{figure}

\begin{figure}
\epsscale{.50} \plotone{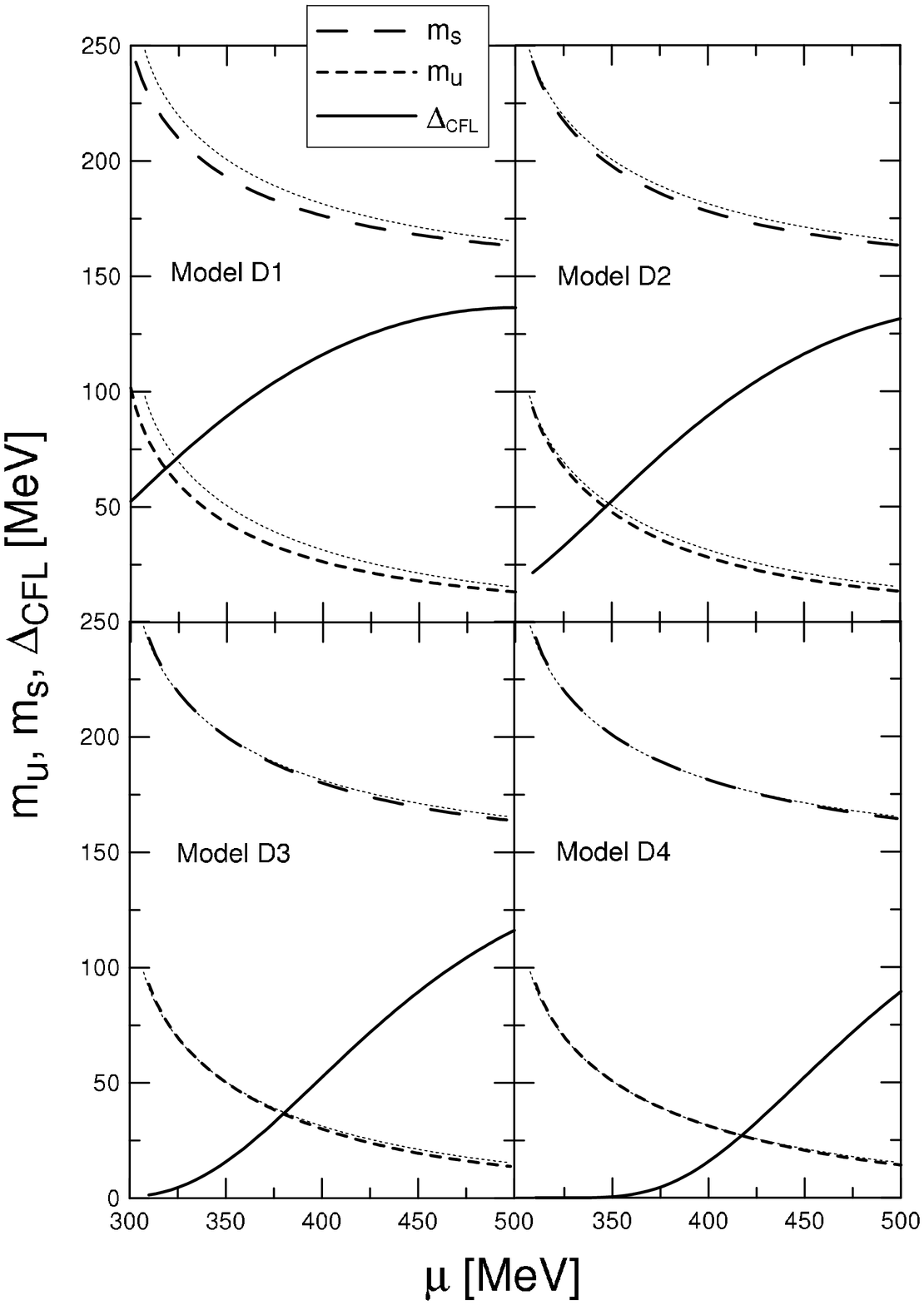} \caption{ Dependence of the
quark masses and the color superconducting gap on the quark
chemical potential for the models D1, D2, D3, and D4 given in
Table 1. The thin dotted curves near the constituent quark masses
curves represent the corresponding constituent quark masses for
the unpaired quark phase. \label{fig2}}
\end{figure}

 The condensation energy of the Cooper pairs in the CFL phase
 given by Eq. (\ref{gapCFL}) \citep{Alford4}, requires the specification of the color
superconducting gap. In order to make the present phenomenological
approach compatible with microscopic treatments based on some
results of QCD, here we use solutions provided by the
Nambu-Jona-Lassinio (NJL) model with a local four-fermion
interaction. A recent study in this direction discussing the
solution of the gap equation is presented in Ref. \citep{Raja},
where it is given the following approximate gap expression:
\begin{equation}
\Delta_{CFL} \simeq 2 \sqrt{\Lambda^2 - \tilde{\mu}^2} \exp \left(
\frac{\Lambda^2 - 3\tilde{\mu}^2}{2 \tilde{\mu}^2} \right) \exp
\left( - \frac{\pi^2}{8 G_{D} \tilde{\mu}^2} \right)
\label{Delta},
\end{equation}
where $G_{D} = \eta G_{S}$ is the strength of the diquark pairing,
with $G_{S}$ being the quark-antiquark coupling constant and
$\eta$ a dimensionless parameter between $3/4$, for the
intermediate coupling strength, and $1$ corresponding to the
strong coupling between quarks \citep{Ruster}. In the above
equation we have introduced a quark chemical potential shift,
$\tilde{\mu} = \mu - {\mu}^{*} $, where ${\mu}^{*}$ establishes
the quark chemical potential value for which the gap begins to
take significant values.

At this point one can realize that the present calculation
involves two parameters from the dynamical quark mass expression
($C$ and $m_{s0}$) and the parameter $\mu^{\ast}$ to adjust the
gap in the density region of interest. The values of the first
pair of parameters are chosen according to the stability criterion
applied to the strange quark matter energy per baryon, which reads
$\epsilon / \rho_{0} \le 930 MeV$ at normal nuclear density
$\rho_{0}$, where $930$ $MeV$ is the energy per baryon in iron
nuclei \citep{Oliveira}. The possible values of the parameter $C$
lie in the interval $69.05\leq C\leq 111.6$ $MeV . fm^{-3}$ and
the strange quark mass lies in the interval $40 \leq m_{s0} \leq
180$ $MeV$ \citep{Lugones}. The current masses of up and down
quarks are set $m_{u0}=m_{d0}=0$. For the free parameter
${\mu}^{*}$ we have used the values  $100$, $150$, $200$, and
$250$ $MeV$ to adjust the gap behavior properly in the range of
densities covered by compact objects in analysis. Different
combinations of the parameters values are considered in this work.
In Table \ref{t1} we display the different sets of the model
parameter values applied in the present study. For the whole
calculation we have fixed $\Lambda = 603 $ $MeV$ and $
G_{S}\Lambda^2 =1.822 $. We will consider only the case
$G_{D}=G_{S}$ for the diquark coupling strength.

In figures \ref{fig1} and \ref{fig2} we show the dependence of the
constituent quark masses and the color superconducting gap as
functions of the quark chemical potential, for the cases $A$ and
$D$ displayed in Table \ref{t1}, covering four values of the free
gap parameter $\mu^{\ast}$. We can see that for $\mu^{\ast} = 100$
$MeV$, the color superconducting phase dominates even for low
values of the quark chemical potential. However, for higher values
of the parameter $\mu^{\ast}$, the unpaired phase dominates for
low values of the quark chemical potential and thus we need to
deal with a phase transition between the two phases. Similar
behavior for the gap and the constituent masses is observed for
the other sets of parameters given in Table \ref{t1}. Consequently
it is necessary to analyze phase transitions between the color
superconducting and unpaired quark matter, as well as how
sensitive is the phase transition to the parameter changes. We
remark that the quark pairing lowers the constituent quark masses
when compared to the values in the unpaired quark phase.

\section{Color superconducting to unpaired quark matter phase transition}

For the unpaired quark matter (UQM), the thermodynamic potential
density reads
\begin{equation}
\Omega_{UQM} =
 \sum_{f}\frac{3}{\pi^{2}}\int_{0}^{\nu_{f}}k^{2}(\sqrt{k^{2}
+m_{f}^{2}}-\mu_f)dk  \label{potential1} ,
\end{equation}
with $f=\{u,d,s\}$ and where $\nu_{f}$ is the Fermi momentum of
each quark flavor, which is given by
\begin{equation}
\nu_f = \sqrt{\mu_f^2 -m_f^2} .
\end{equation}

The number density of each quark flavor is defined by
\begin{equation}
n_{f}= -\left.
\frac{\partial\Omega_{UQM}}{\partial\mu_{f}}\right\vert_{\rho_{B}},
\end{equation}
which provides explicitly
\begin{equation}
n_{f} = \frac{1}{\pi^2} (\mu_{f}^2 - m_{f}^2)^{3/2} .
\end{equation}
For degenerate and massless free electrons, the free  energy
density is given by
\begin{equation}
\Omega_{e} = - \frac{1}{12 \pi^{2}}{\mu_e}^4 \label{potentiale} ,
\end{equation}
and thus the electron number density reads
\begin{equation}
n_{e} = \frac{1}{3\pi^2} \mu_{e}^3 .
\end{equation}

 In beta equilibrium, the chemical potentials of
quarks and electrons must satisfy the following relationships:
\begin{equation}
\mu_{d} = \mu_{u} + \mu_{e}, \, \, \, \, \, \, \, \,
\mu_{s}=\mu_{d}    \label{chemicalpotential} .
\end{equation}

The condition of electric charge neutrality provides
the equation
\begin{equation}
\frac{2}{3} n_{u} -\frac{1}{3} n_{d} -\frac{1}{3} n_{s} - n_{e} =
0 \label{charge}.
\end{equation}

    Note that the existence of different quark
masses turns impossible for the unpaired quark matter to satisfy
equations (\ref{chemicalpotential} ) and (\ref{charge})
simultaneously, without the presence of electrons in bulk.

\begin{figure}
\epsscale{.70} \plotone{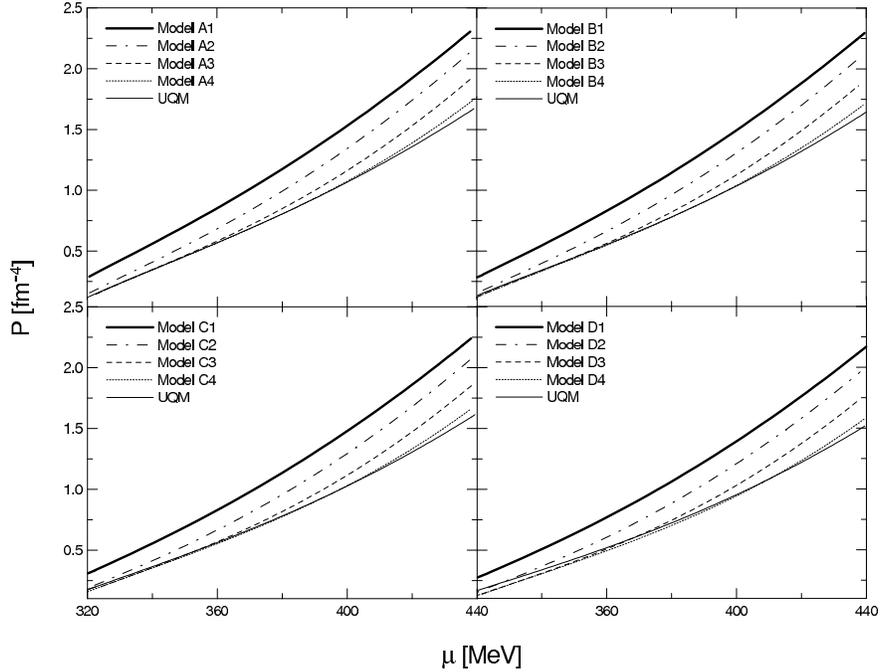} \caption{ Pressure for
different values of the model parameters given in Table \ref{t1},
and for the unpaired quark matter phase as functions of the quark
chemical potential.  \label{fig3}}
\end{figure}

 In order to analyze the occurrence of the phase transition, the
behavior of the quark matter pressure as a function of the quark
chemical potential is depicted in the Fig. \ref{fig3} for all sets
of parameters combination given in Table \ref{t1}. For the class A
models, we can identify subtle intersections of the pressure
curves for the cases A3 and A4 with the curve of the unpaired
phase, which indicate the phase transitions between the two
phases, occurring at $\mu \simeq 335 $ $MeV$ (case A3) and at $\mu
\simeq 390$ $MeV$ (case A4). For the set of parameters D2, D3 and
D4 the phase transition between the two phases occurs at $\mu
\simeq 333 $ $MeV$ (case D2), $\mu \simeq 370 $ $MeV$ (case D3),
and at $\mu \simeq 413 $ $MeV$ (case D4). For the last one, the
phase transition occurs at a baryon density $\rho \simeq 5.5
\rho_0$, with $\rho_0 = 0.15 $ $fm^{-3}$ being the normal nuclear
matter density. We remark that small values of the parameter
$\mu^{\ast}$ combined with the high values of the current strange
quark mass provide stiffer equations of state, leading to more
massive quark stars, as we will discuss latter.

\section{Quark star structure}

The calculation of the structure of the quark star models for
given values of the central density is giving by solving
numerically the Tollman-Oppenheimer-Volkoff (TOV) equations
\citep{TOV2,TOV1}:
\begin{equation}
\frac{dP}{dr}=-G\frac{m(r)\varepsilon (P)}{(rc)^{2}}\left( 1+\frac{P}{%
\varepsilon }\right) \left( 1+\frac{4\pi r^{3}P}{m(r)c^{2}}\right)
\left( 1-2G\frac{m(r)}{rc^{2}}\right) ^{-1},
\end{equation}
and
\begin{equation}
\frac{dm(r)}{dr}=4\pi r^{2}\varepsilon.
\end{equation}

 The obtained results show that larger values of
$\Delta_{CFL} $ and the current strange quark mass tend to give
more massive and larger quark stars. With an appropriate choice of
the gap parameter $\mu^{\ast}$, it is possible to obtain masses
and radii compatible with the neutron stars calculated with
equation of states widely used to nuclear and hadronic matter.
Consequently, suitable selection of quark matter equation of state
may lead to similar observational results obtained in different
constitutive theoretical frameworks to the dense stellar matter
composition.

\begin{figure}
\epsscale{.60} \plotone{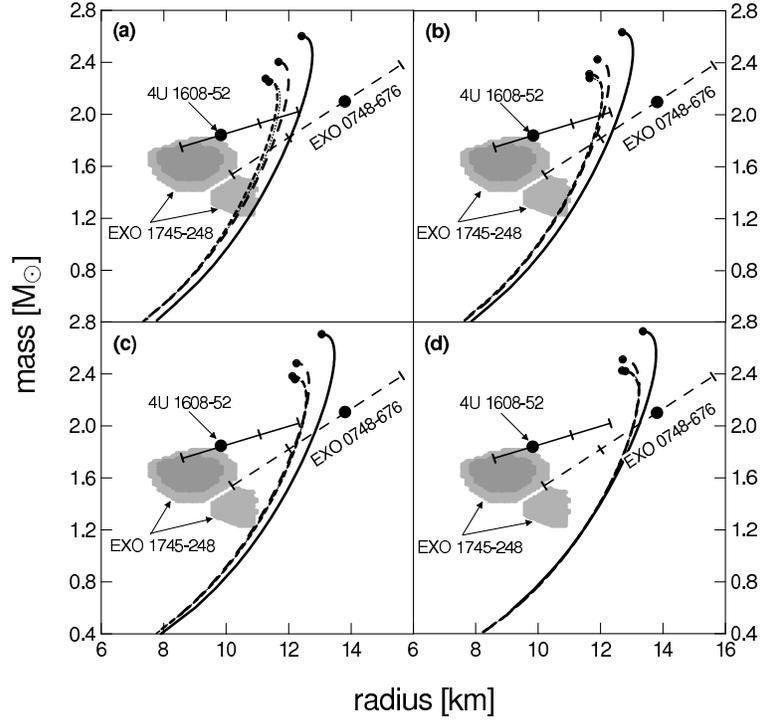} \caption{ Mass-radius
diagrams for the stable quark stars. The cases A1 to A4 given in
Table \ref{t1} are shown in part-a; the cases B1 to B4, in part-b;
the cases C1 to C4, in part-c; and the cases D1 to D4 in part-d.
In all parts of the panel the solid lines stand for the models A1,
B1, C1, and D1 ($ \mu^{\ast} = 100$ $MeV$); the long-dashed curves
are for the models A2, B2, C2, and D2 ($ \mu^{\ast} = 150$ $MeV$);
the short-dashed curves represent the models A3, B3, C3, and D3 ($
\mu^{\ast} = 200$ $MeV$); and the thin dotted curves represent the
models A4, B4, C4, and D4 ($ \mu^{\ast} = 250$ $MeV$). It is also
shown the lower limits on the mass and radius of the neutron star
$EXO$ $0748-676$, given by $M=2.10 \pm 0.28$ $M_{\odot}$ and
$R=13.8\pm 1.8$ $km$. The one- and two-$\protect\sigma $ error
bars of the results obtained by \"{O}zel are also shown (dashed
bars). Regarding the neutron star in $EXO$\ $1745-248$, \"{O}zel
have found two pairs of values for the mass and radius, which are
centered around $M=1.4\ M_{\odot }$ and $R=11\ km$\ or around
$M=1.7\ M_{\odot }$ and $R=9\ km$. The one- and two-
$\protect\sigma $ contour lines for its mass and radius are
represented by the two gray regions. For the neutron star in $4U\
1608-52$ the obtained values are $M=1.84\pm 0.09\ M_{\odot }$ and
$R=9.83\pm 1.24\ km$. The one- and two-$\protect\sigma $ error
bars of the new results obtained by \"{O}zel are shown (solid
bars). For all error bars in the figure the one-$\protect\sigma $
are in both directions and the two-$\protect\sigma $ are only in
one direction.\label{fig4}}
\end{figure}

In the panel showed in the Fig. \ref{fig4} we compare the limits
of mass and radius given in the \"{O}zel's works with those of
quark stars calculated in this work, considering only the regime
of strong coupling with $G_{D} = G_{S}$. Our results have been
obtained for all sets of values of the parameters given in Table
\ref{t1}. The corresponding quark star models can reach a mass of
$2.25$ $M_{\sun}$ and radius $11.4$ $km$, for $\mu^\ast = 250$
$MeV$, as showed in part-a of the panel, up to the maximum value
$2.73$ $M_{\sun}$ with the radius $13.3$ $km$, for $\mu^\ast =
100$ $MeV$, as showed in part-d of the panel. For the sake of
comparison, we show in the same panel the lower limits on the mass
and radius of the neutron star EXO 0748-676, with the values of
mass and radius given by $M=2.10 \pm 0.28$ $M_{\sun}$ and $R=13.8
\pm 1.8$ $km $. The one- and two-$\sigma$ error bars of the
results obtained by \"{O}zel are shown (dashed bars). \"{O}zel
have found two pairs of values for the mass and radius of the
neutron star in EXO 1745-248, which are centered around $M=1.4$
$M_{\odot}$ and $R=11$ $km $ or around $M=1.7$ $M_{\sun}$ and
$R=9$ $km$. The one- and two-$\sigma$ contours for its mass and
radius are represented by the dark and soft grey regions,
respectively. For the neutron star in 4U 1608-52 the obtained
values are $M=1.84 \pm 0.09$ $M_{\odot}$ and $R=9.83 \pm 1.24$
$km$.

We can see that the obtained values of masses and radii for pure
strange quark star models are in good agreement with those given
by \"{O}zel in the references \citep{Ozel1} and \citep{Ozel3}, and
only in fair accordance with present data obtained for the compact
star $EXO$ $1745-248$, which corresponds to the smallest radius of
the set of compact stars showed in the figure. From the result
obtained for this class of object we thus can infer that such type
of small compact stars, with small radii, are not natural
candidates for pure quark stars, as it happens for the other ones,
more massive and with lager radii. Nevertheless, small compact
stars may be identified with hybrid stars composed of a quark
matter core surrounded by a hadronic envelope. The present quark
matter equation of state could still be helpful to describe the
quark matter in the core of these hybrid structures.

\section{ Quark star with a nuclear crust}

 The strange quark matter in CFL phase is electrically neutral
since the Cooper quark pairing minimizes the energy if the quarks
have equal Fermi momenta. In this situation, the number of $u$,
$d$ and $s$ quarks are equal and the bulk electric charge is zero
without the need of electrons in the medium. However, even for
strange quark matter in the CFL phase, when surface effects are
taken into account, the number of massive quarks is reduced near
the boundary of the star relative to the number of massless quarks
at fixed Fermi momenta values. Consequently, there is a net quark
charge at the very thin surface of the system. The thickness of
the charged surface layer of the CFL quark star is $\sim 1$ $fm$,
as it was discussed in references \citep{Jaffe,Madsen,Usov}. The
presence of such strong electric fields near the surface of the
CFL strange star, make possible the existence of nuclear crusts
spatially separated from the deconfined CFL quark core
\citep{Fridolin}.

Therefore, in order to investigate the change in the quark star
structure with the inclusion of a nuclear matter crust, we have
constructed an equation of state for the crust assuming that cold
stellar matter in this crust is composed of a nuclear lattice
immersed in an electron gas. To describe this crust matter we have
used the equation of state given in \citep{Baym}. Hence, we can
find equilibrium configurations compatible with the solution of
the TOV equations using this equations of state connected to the
CFL matter equation of state previously constructed.

\begin{figure}[th]
\vspace*{.5 cm} \centerline{
\resizebox{8.cm}{!}{\includegraphics{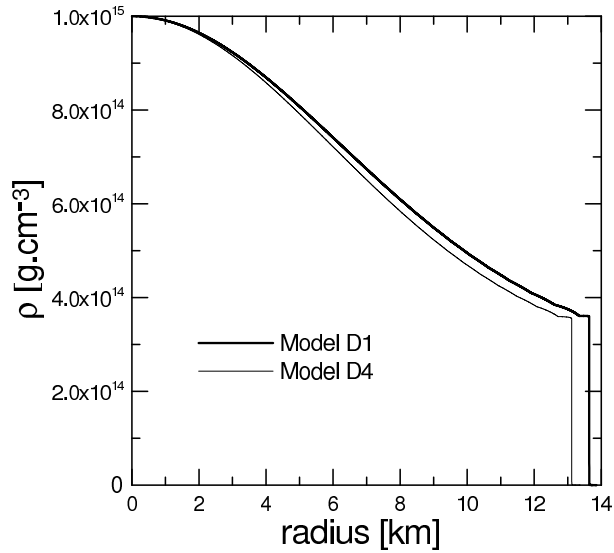}}}
\vspace*{.25cm} \caption{Energy density versus radius of quark
stars with central density $\rho_{c}=1.0 \times 10^{15}$ $g
cm^{-3}$ for the models D1 and D4 given in Table
\ref{t1}.\label{fig5}}
\end{figure}

\begin{figure}[th]
\vspace*{.5 cm} \centerline{
\resizebox{8.cm}{!}{\includegraphics{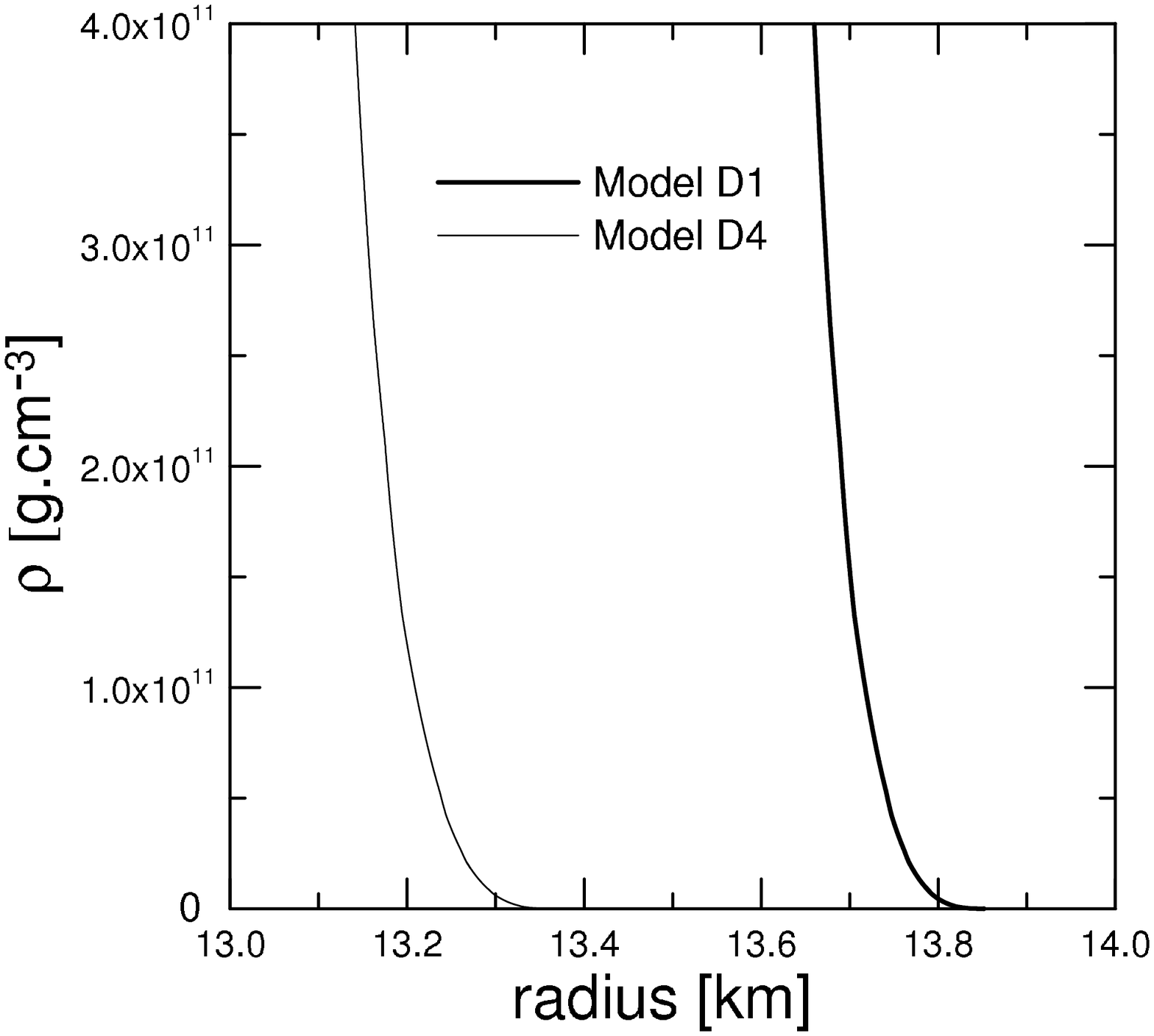}}}
\vspace*{.25cm} \caption{Details of the nuclear crusts of the
quark stars displayed in the last figure.\label{fig6}}
\end{figure}

\begin{figure}[th]
\vspace*{.5 cm} \centerline{
\resizebox{8.cm}{!}{\includegraphics{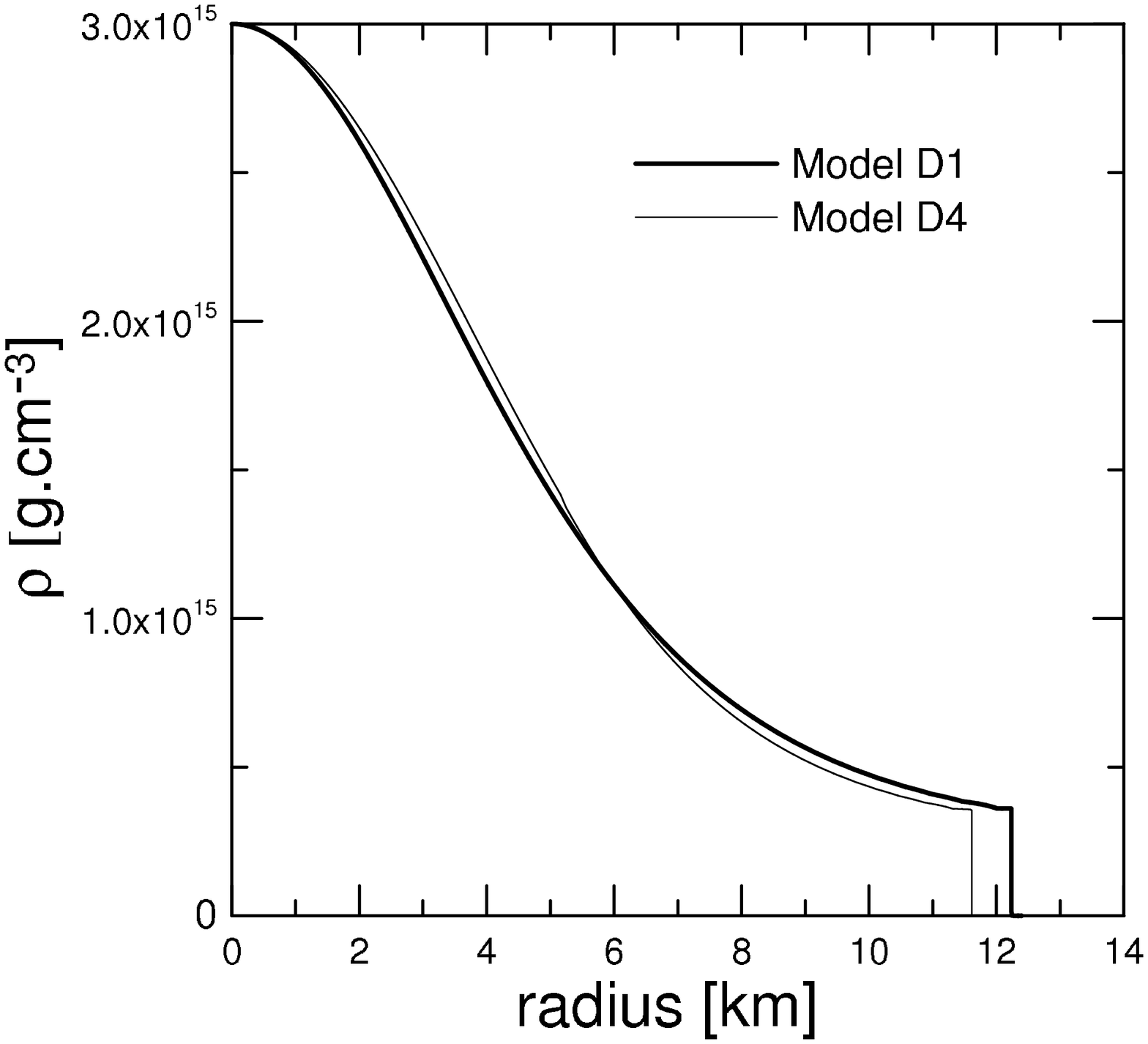}}}
\vspace*{.25cm} \caption{Energy density versus radius of quark
stars with central density $\rho_{c}=3.0\times 10^{15}$ $g
cm^{-3}$ for the models D1 and D4 given in Table
\ref{t1}.\label{fig7}}
\end{figure}

\begin{figure}[th]
\vspace*{.5 cm} \centerline{
\resizebox{8.cm}{!}{\includegraphics{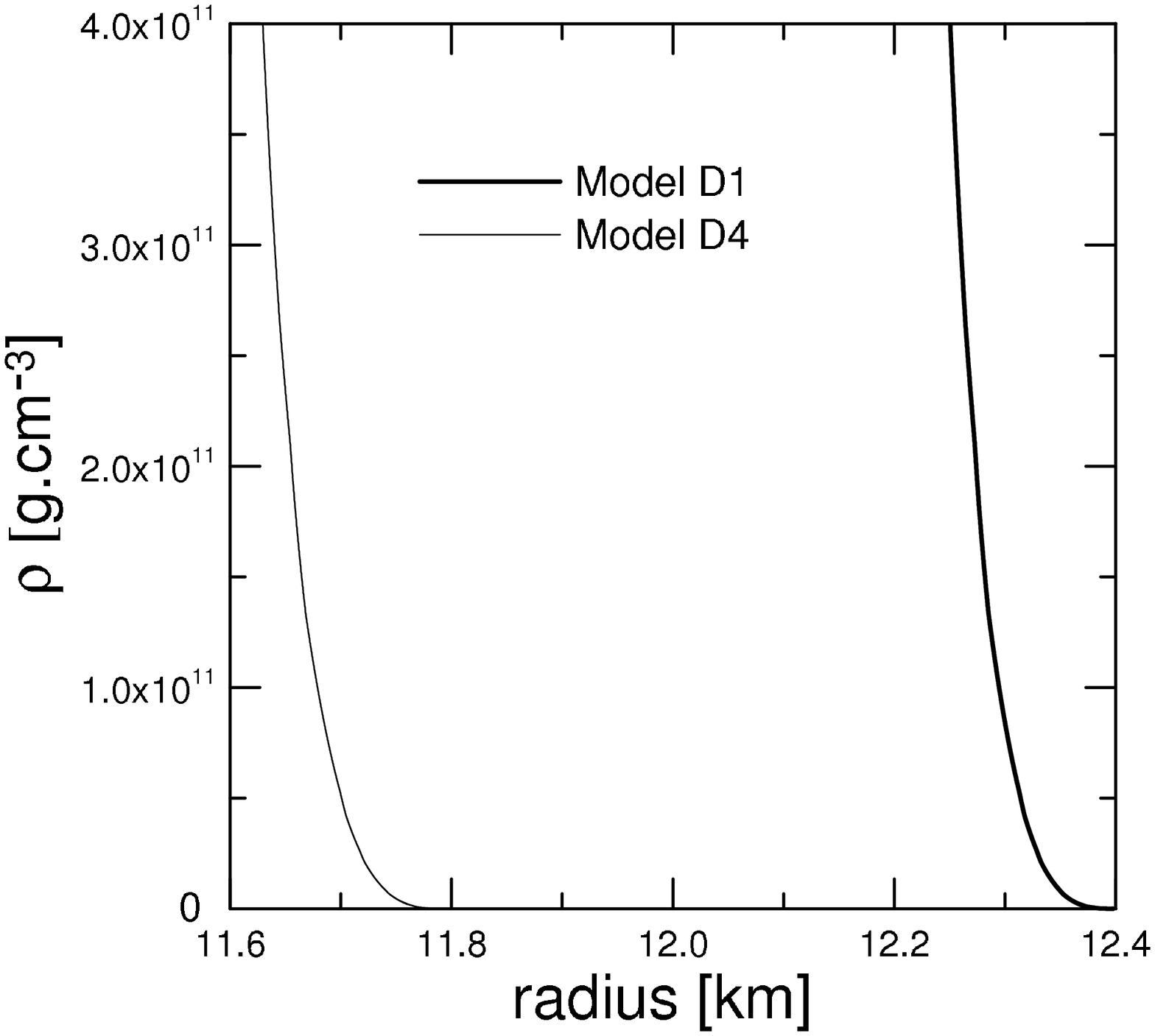}}}
\vspace*{.25cm} \caption{Details of the nuclear crusts of the
quark stars displayed in the last figure.\label{fig8}}
\end{figure}

In Fig. \ref{fig5} we depict the energy density as a function of
radius for two quark star models with the central density $1.0
\times 10^{15}$ $g cm^{-3}$, for the set of parameters D1 and D4
of Table \ref{t1}. In the figure are depicted the density as a
function of radius for two compact stars with dense and large
cores. That one corresponding to the set of parameters D4
($\mu^{\ast} = 100$ $MeV$) is composed entirely of deconfined
quarks in the $CFL$ phase, and the other one, for the case A1
($\mu^{\ast} = 250$ $MeV$) is made of unpaired quarks. The
obtained masses for these two compact stars are $2.69$ $M_{\sun}$,
for the first model, and $2.37 \ M_{\sun} $, for the second one.

 Figure \ref{fig6} shows the detailed structure of the nuclear
crusts enveloping the quark matter cores of the model compact
stars shown in the previous figure. The thickness of these nuclear
crusts are $196$ $m $, for the model D1 ($\mu^{\ast} = 100$
$MeV$), and $231$ $m$ for the model D4 ($\mu^{\ast} = 250$ $MeV$).

 In Fig. \ref{fig7} are displayed the energy density as a
function of radius for two quark stars with the central density
$3.0 \times 10^{15}$ $g cm^{-3}$, again using the set of
parameters of the models D1 and D4. For the model D1 the obtained
quark star has a large core composed entirely of deconfined quarks
in the $CFL$ phase, and for the model D4 the large and dense core
is composed of CFL quark phase enveloped by an unpaired quark
shell. The obtained masses are $2.56$ $M_{\sun}$, for the the case
D1, and $2.28$ $M_{\sun}$, for the case D4. The detailed
structures of the nuclear crusts enveloping the quark stars shown
in the previous figure are displayed in Fig. \ref{fig8}. The
thickness of these nuclear crusts are $150$ $m $, for the first
one, and $169$ $m$, for the second one.

The obtained results indicate that smaller values of $\mu^{\ast}$
correspond to more massive and less compact quark stars, almost
entirely composed of quark matter. An interesting implication of
the obtained results is that for increasing $CFL$ gap energies one
obtains smaller nuclear crusts. However, the $CFL$\ quark star
will be naked only for a very unrealistic high value of the $CFL$
gap energy.

Observe that the maximum density at the base of the nuclear crust
which envelopes the quark star is determined by the threshold
density for neutron drip ($\rho \sim 4.0\times 10^{11}$ $g
cm^{-3}$). On the other hand, the equal values of pressure at the
surface of the quark matter core and the base of the nuclear crust
is necessary to assure the hydrostatic equilibrium of the quark
star as a whole.

\section{Conclusions}

In this work we have studied the properties of quark matter
applying the density-dependent quark mass model in the calculation
of the equation of state of the color superconducting quark
matter. Within this framework, we have investigated the quark star
structure for suitable values of the model parameter $C$, the
current mass of the quark $s$, and the $CFL$ gap energy, and so we
have discussed the dependence of the observable properties of
quark stars, such as maximum mass and radius, on the $CFL$ gap
energy.

In the present phenomenological analysis, we have used a color
superconducting gap parametrization derived from microscopic
approaches widely applied to describe color superconducting quark
matter. The density-dependent quark mass model is also used in
order to mimic the quark chiral symmetry behavior.

 According to our results, the existence of the reported massive compact stars
can be compatible with stable quark stars with a nuclear crust, or
even bare quark stars.

The results presented here show that massive compact stars with
mass $ \gtrsim 2 M_{\sun}$ up to maximal masses $2.73$ $M_{\sun}$
would be interpreted as compact stars composed entirely of
deconfined quarks, in the color superconducting phase or even in
the unpaired phase. Moreover, less massive and smaller compact
stars would be identified with hybrid stars composed of a
deconfined quark phase in the inner core enveloped by a hadronic
phase, spatially separated due to the gravitational action. This
configuration is allowed since the burning of nuclear matter into
quarks must be an exothermic process. This specific situation can
occurs for a range of densities, which are determined by the
equations of state used to describe both phase, as shown in Refs.
\citep{Lugones,Lugones2}.

 As a final remark, we call attention that the observational data
obtained for compact objects which are possible candidates to be
quark stars could be used to constrain the parameters of
phenomenological models involved in the quark matter equation
construction.

\acknowledgments

The authors are grateful to F. Weber for valuable discussions. The
authors would like to thank the referee for the fruitful and
clarifying discussions during the preparation and revision of the
manuscript. H. Rodrigues and S. B. Duarte thank CNPq for financial
support.


\end{document}